\documentstyle[epsf,12pt]{article}
\begin{document}

\thispagestyle{empty}
\onecolumn
\date{\today}
\vspace{-1.4cm}
\begin{flushleft}
{BUTP-99/11\\}
{BI-TP 99/08\\}
{DESY 99-086 \\}

\end{flushleft}
\vspace{0.3cm}

\begin{center}

{\LARGE {\bf
Algebraic reduction of one-loop Feynman graph amplitudes
 }\vglue 10mm
        }


\vfill
{\large
 J. Fleischer$^{a}$, F. Jegerlehner$^{c}$ and
  O.V. Tarasov$^{a,b,c,}$\footnote{On leave of absence from JINR,
141980 Dubna (Moscow Region), Russian Federation.}$^,\!\!$
\footnote{
Supported by BMBF
under contract PH/05-7BI92P 9
}
}

\vspace{1cm}
$^a$~~Fakult\"at f\"ur Physik~~~~~\\
Universit\"at Bielefeld \\
Universit\"atsstr. 25\\
D-33615 Bielefeld, Germany\\
\vspace{0.3cm}

$^b$~~Institute of Theoretical Physics,\\
University of Bern,\\
Sidlerstrasse 5, CH-3012 Bern, Switzerland

\vspace{0.3cm}

$^c$~~Deutsches Elektronen-Synchrotron DESY~~~~~ \\
Platanenallee 6, D--15738 Zeuthen, Germany\\
\end{center}

\vfill

\begin{abstract}
An algorithm for the reduction of one-loop $n$-point tensor
integrals to basic integrals is proposed.
We transform tensor integrals to scalar integrals with shifted
dimension \cite{Andrey,ArBoos} and reduce these by recurrence relations 
to integrals in generic dimension \cite{ovt1}. 
Also the integration-by-parts method \cite{ibpm} 
is used to reduce indices (powers of scalar propagators) 
of the scalar diagrams.
The obtained recurrence relations for one-loop integrals
are explicitly evaluated for 5- and 6-point functions.
In the latter case the corresponding Gram determinant vanishes
identically for $d=4$, which greatly simplifies the application of the
recurrence relations. 
\end{abstract}

\newpage

\section{Introduction}
Radiative corrections contain not only most essential
information about quantum properties of a quantum field theory
but moreover, their knowledge is indispensable for the
interpretation of precision experiments. Higher order calculations in
general have been performed for processes with at most four external
legs. With increasing energy the observation of multi particle events
is becoming more and more important and at least one-loop diagrams
must be calculated for these cases as well. One of the important
examples is the $W$--pair production reaction $e^+ e^- \rightarrow
W^+W^- \rightarrow 4\;\mathrm{fermions}\;(\gamma)$, being
experimentally investigated at LEP2, which allows us to accurately
determine mass, width and couplings of the $W$-boson. At future
high--energy and high--luminosity $e^+e^-$ linear colliders radiative
corrections will become even more important for a detailed
understanding of Standard Model processes. Thus their computation will
be crucial for a precise investigation of particle properties and the
expected discovery of new physics. At present, a full $O(\alpha)$
calculation for a four--fermion production process is not available,
mainly because, such a calculation is complicated for various
reasons~\cite{DDRW,Vici}. One of the problems is addressed in the
present article and concerns the calculation of one-loop diagrams with
special emphasis on processes with five and six external legs.

The evaluation of one-loop Feynman graph amplitudes
has a long history \cite{brown} - \cite{Wein}
and at the present time many different methods and
approaches do exist. However, the most frequently used
Passarino-Veltman \cite{pv} approach is rather difficult to use
in calculating diagrams with five, six or more external legs.
The tensor structure of such  diagrams  is rather
complicated. To obtain the coefficients in the tensor decomposition
of multi-leg integrals one needs to solve 
algebraic equations of high order which often are not
tractable even with the help of modern computer algebra
systems. Additional complications occur in the case
when some kinematic Gram determinants are zero.

 In the present paper we propose an approach which 
simplifies the evaluation of one-loop diagrams in an essential way. 
It allows one to evaluate multi-leg integrals efficiently
without solving systems of algebraic equations.
Integrals with five and six external legs are worked out 
explicitly. Those with seven and more external legs
need special investigation because there are different types
of Gram determinants vanishing and their consideration is postponed
for this reason. The above cases with more than four
external legs allow particular simplifications: for n=5 due to 
the property (\ref{siebzehn}) and the property of the recurrence 
relation (\ref{sechsundzwanzig}) and for $n > 5$ due to properties 
of Gram determinants.
Our scheme allows also to evaluate diagrams with $n \leq 4$, but
here we cannot use the above mentioned properties. A simplification
is achieved in this case since solving of linear equations is avoided
and the algorithm is implemented in a FORM \cite{FORM} program for
these cases.

\section{Recurrence relations for \lowercase{n}-point 
 one-loop integrals}

First we consider  scalar one-loop integrals  depending
on $n-1$  independent external momenta:
\begin{equation}
I_n^{(d)}=
\int \frac{d^d q}{\pi^{{d}/{2}}} \prod_{j=1}^{n}
\frac{1}{c_j^{\nu_j}},
\end{equation}
where
\begin{equation}
c_j=(q-p_j)^2-m_j^2+i\epsilon ~~~{\rm for}~~{j<n}~~~{\rm and}
~~~c_n=q^2-m^2_n+i\epsilon.
\end{equation}
The corresponding diagram and the convention for
the momenta are given in Fig.1.


%
%
%

\begin{center}
\vspace*{-2cm}
\vbox{
 \raisebox{5.0cm}{\makebox[0pt]{\hspace*{-2cm}$$}}
 \epsfysize=30mm \epsfbox{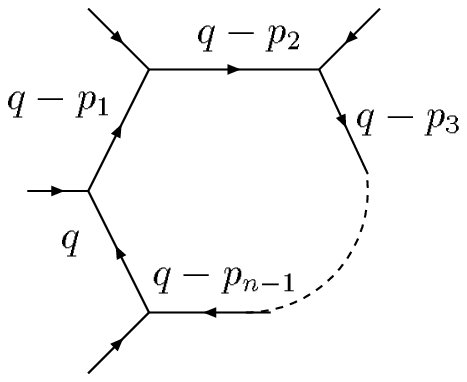}
     }
\end{center}
\vspace*{0.7cm}
\noindent
\begin{center}
Fig. 1: One-loop diagram with $n$ external legs.
\end{center}

Tensor integrals 
\begin{equation}
I_{n,r}^{(d)}=
\int \frac{d^d q}{\pi^{{d}/{2}}} 
\prod_{j=1}^{n}\frac{q_{\mu_1} \ldots q_{\mu_r}}{c_j^{\nu_j}},
\end{equation}
can be written as a combination of scalar integrals
with shifted space-time dimension multiplied by tensor structures
in terms of external momenta and the metric tensor. This was shown in
Ref.\cite{Andrey} for the one-loop case and in Ref.\cite{ovt1}
for an arbitrary case.  We shall use the relation proposed in \cite{ovt1} 

\begin{equation}
I_{n,r}^{(d)}=T_{\mu_1 \ldots \mu_r} (\{ p_s \},\{ \partial_j \}, {\bf d^+})
I_{n}^{(d)},
\label{tensint}
\end{equation}
where $T$ is a tensor operator, $\partial_j=\frac{\partial}{\partial m_j^2}$,
and ${\bf d^+}$ is the operator shifting the value of the space-time
dimension of the integral by two units: ${\bf d^+}I^{(d)}=I^{(d+2)}$.
On the right-hand side of (\ref{tensint}) it is assumed that the
invariants $c_i$ have arbitrary masses and only
after differentiation with respect to $m_i^2$ these are set to
their concrete values.

To derive an explicit expression for the tensor operator
$T_{\mu_1\ldots \mu_r} (\{p_s\},\{{\partial}_j\},{\bf d^+}) $
we introduce an auxiliary vector $a$  and write the tensor structure of 
the integrand as
\begin{equation}
q_{\mu_1} \ldots q_{\mu_r} =
\left.\frac{1}{i^{r}}\frac{\partial}{\partial a_{\mu_1}}
 \ldots \frac{\partial}{\partial a_{\mu_r}}
 \exp \left[i a q\right]
\right|_{ a_i=0 }.
\label{aderiv}
\end{equation}
Next we transform the integral
\begin{equation}
I^{(d)}_n(a)=\int \! \frac{d^d q}{\pi^{{d}/{2}}} \prod_{j=1}^{n}
\frac{1}{c_j^{\nu_j}}
\exp \left[i (a q) \right];~~~I^{(d)}_n \equiv I^{(d)}_n(0).
\end{equation}
into the $\alpha$-parametric representation by means of
\begin{equation}
\frac{1}{(k^2-m^2+i\epsilon)^{\nu}}
 = \frac{i^{-\nu}}{ \Gamma(\nu)}\int_0^{\infty}
 d\alpha ~\alpha^{\nu-1} \exp\left[i\alpha(k^2-m^2+i\epsilon)\right],
\end{equation}
and perform the $d$-dimensional Gaussian integration
\begin{equation}
\int d^dk \exp \left[i(A k^2+ 2(pk))\right] =i
 \left( \frac{\pi}{i A} \right)^{\frac{d}{2}}
 \exp \left[ -\frac{ip^2}{A} \right] .
\end{equation}
The final result is:
\begin{equation}
I^{(d)}_n(a)\!=i\!\left ( \frac{1}{i} \right)^{d/2}
\! \prod^{n}_{j=1} \frac{i^{-\nu_j}}{\Gamma(\nu_j)}
\! \int_0^{\infty} \!\!\!\! \ldots \!\! \int_0^{\infty}
\frac{d \alpha_j \alpha^{\nu_j-1}_j}
     { [ D(\alpha) ]^{\frac{d}{2}}}
  \exp \left[
              i \left(\frac{Q(\alpha,\{ p_s \},a)}
                           {D(\alpha)}
             \!-\!\sum_{l=1}^{n}\alpha_l(m_l^2\!-\!i\epsilon)
                \right)
         \right],
\label{repres}
\end{equation}
where
\begin{eqnarray}
\label{Dform}
&&D(\alpha)=\sum_{j=1}^{n} \alpha_j, \\
&& \nonumber \\
&&Q(\alpha,\{p_s\},a)=Q(\alpha,\{p_s\})+\sum_{k=1}^{n-1}
 (ap_k) \alpha_k -\frac14 a^2
\end{eqnarray}
and $Q(\alpha,\{p_s\})$ is the usual $a$-independent $Q$-form of the graph.
From the representation (\ref{repres}) together with (\ref{aderiv}) it is 
straightforward to work out that
\begin{equation}
T_{\mu_1 \ldots \mu_r}(\{ p_s \},\{ \partial_j \}, {\bf d^+})
=\frac{1}{i^r} \prod_{j=1}^r \frac{\partial}{\partial a_{\mu_j}}
 \exp \left[i \left( \sum_{k=1}^{n-1}(ap_k) \alpha_k-\frac14 a^2 \right )
 \rho
    \right] 
 \left|_{ a_j=0 \atop {\alpha_j=i \partial_j \atop
  \rho=i {\bf d^+}} } \right. .
\label{Ttensor}
\end{equation}
This representation is particularly well suited and effective for a computer 
implementation of the tensor integrals (\ref{tensint}).


\subsection{Integrals with non zero Gram determinants}

The purpose of this Sect. is to develop
an algorithm for reducing the above 
mentioned scalar integrals to standard integrals in generic dimension
$d=4-2 \varepsilon$.

Recurrence relations which reduce the index of the $j$-th line
without changing the space-time dimension are obtained by the
integration-by-parts method \cite{ibpm}:

%
%
\begin{eqnarray}
&&2 \Delta_n \nu_j {\bf j^+} I_n^{(d)}=
  \sum^{n}_{k=1} (1+\delta_{jk})   \left( \frac{\partial \Delta_n}
 {\partial Y_{jk}} \right)
\left[ d - \sum_{i=1}^{n} \nu_i( {\bf k^-} {\bf i^+}+1)
             \right] I_n^{(d)},
\label{recurseJ}
\end{eqnarray}
where $\delta_{ij}$ is the Kronecker delta symbol, 
the operators ${\bf j^{\pm }}$ etc.
shift the indices $\nu_j \to \nu_{j } \pm 1$ and
$$
\Delta_n=  \left|
\begin{array}{cccc}
Y_{11}  & Y_{12}  &\ldots & Y_{1n} \\
Y_{12}  & Y_{22}  &\ldots & Y_{2n} \\
\vdots  & \vdots  &\ddots & \vdots \\
Y_{1n}  & Y_{2n}  &\ldots & Y_{nn}
\end{array}
         \right|.
$$
Taking derivatives of $\Delta_n$ one should
consider all $Y_{ij}$  as independent
variables and for $j>k$ assume $\partial / \partial Y_{jk}
= \partial / \partial Y_{kj}$. After taking derivatives one should set
\begin{equation}
Y_{ij}=-(p_i-p_j)^2+m_i^2+m_j^2,
\end{equation}
where $p_i,p_j$ are external momenta flowing through $i-,j$-th 
lines, respectively, and $m_j$ is the mass of the $j$-th line ($p_n=0$).
At this stage the external momenta are not yet restricted to dimension 4.
We will specify later when this property is used.

A recurrence relation for reducing dimension and index of
the $j$-th line is obtained from \cite{ovt1}:

\begin{equation}
G_{n-1} \nu_j{\bf j^+} I^{(d+2)}_n=
\left[ (\partial_j  \Delta_n) +\sum_{k=1}^{n}
 (\partial_j \partial_k \Delta_n)
 {\bf k^-} \right] I^{(d)}_n,
\label{reduceJandDtod}
\end{equation}
where $\partial_j \equiv \partial / \partial m_j^2$ and
\begin{equation}
G_{n-1}= -2^n \left|
\begin{array}{cccc}
  p_1p_1  & p_1p_2  &\ldots & p_1p_{n-1} \\
  p_1p_2  & p_2p_2  &\ldots & p_2p_{n-1} \\
  \vdots  & \vdots  &\ddots & \vdots \\
  p_1p_{n-1}  & p_2p_{n-1}  &\ldots & p_{n-1}p_{n-1}
\end{array}
\right|.
\label{Gn}
\end{equation}

We may also reduce the space-time dimension of $I^{(d)}_n$
by means of:
\begin{equation}
  (d-\sum_{i=1}^{n}\nu_i+1)G_{n-1}I^{(d+2)}_n=
  \left[2 \Delta_n+\sum_{k=1}^n (\partial_k \Delta_n) {\bf k^-}
  \right]I^{(d)}_n.
  \label{reduceDtod}
\end{equation}

Equations (\ref{recurseJ}), (\ref{reduceJandDtod}) and (\ref{reduceDtod})
are our starting point. Some simplifications of these equations can still
be obtained. In particular
we give in the following a compact representation for the mass-derivatives of $\Delta_n$.
First of all we mention the following useful relation between $G_{n-1}$ and $\Delta_n$:
\begin{equation}
  \sum_{j=1}^n \partial_j \Delta_n= - G_{n-1}.
\end{equation}
The basic object for the purpose of expressing the derivatives in 
(\ref{recurseJ}), (\ref{reduceJandDtod}) and (\ref{reduceDtod})
turns out to be the ``modified Cayley determinant'' of the diagram 
with internal lines 1 $\ldots  n $ \cite{melrose}, namely 
\begin{equation}
()_n ~\equiv~  \left|
\begin{array}{ccccc}
  0 & 1       & 1       &\ldots & 1      \\
  1 & Y_{11}  & Y_{12}  &\ldots & Y_{1n} \\
  1 & Y_{12}  & Y_{22}  &\ldots & Y_{2n} \\
  \vdots  & \vdots  & \vdots  &\ddots & \vdots \\
  1 & Y_{1n}  & Y_{2n}  &\ldots & Y_{nn}
\end{array}
\right|,
\label{MCD}
\end{equation}
labeling elements 0 $\ldots n$. ``Signed minors''
\begin{equation}
{j_1  j_2  \ldots \choose k_1  k_2  \ldots }_n
\end{equation}
will be labeled
by the rows $j_1 , j_2 , \ldots$ and columns $k_1 , k_2 , \ldots$ excluded from ${()}_n$ \cite{melrose,Regge}.
E.g. we have
\begin{equation}
\Delta_n = { 0 \choose 0 }_n
\end{equation}
Further relations are
\begin{equation}
\partial \Delta_n / \partial Y_{jk} = (2-\delta_{jk})~{ 0 j \choose 0 k}_n
\end{equation}
\begin{equation}
{\partial }_j \Delta_n =-2 ~{ j \choose 0}_n
\label{arbam}
\end{equation}
\begin{equation}
{\partial }_j {\partial }_k \Delta_n = ~  2 ~ { j \choose k}_n
\end{equation}
and for $p_n=0$ we have
\begin{equation}
G_{n-1}~=~2 ~ {\left( \right)}_n .
\end{equation}
One obvious advantage of the above relations is that they 
can be easily used for numerical evaluation. Recursion (\ref{recurseJ}) then reads
\begin{equation}
{0\choose 0}_n \nu_j {\bf j^+} I_n^{(d)}=
  \sum^{n}_{k=1} {0j\choose 0k}_n 
\left[ d - \sum_{i=1}^{n} \nu_i( {\bf k^-} {\bf i^+}+1)
             \right] I_n^{(d)}.  
\end{equation}
\\
Using \cite{melrose}
\begin{equation}
\sum^n_{k=1} {0 j \choose 0 k}_n =- {0 \choose j}_n ~,
\label{vierundzwanzig}
\end{equation}
we write it in the most convenient form for further evaluation as
\begin{equation}
{0\choose 0}_n \nu_j {\bf j^+} I_n^{(d)}=
\left\{\left(1+\sum^n_{i=1} \nu_i -d \right) {0 \choose j}_n
-\sum^n_{k=1} {0j\choose 0k}_n (\nu_k -1) \right\} I_n^{(d)}
- \sum^n_{i, k\atop i\not= k} {0j\choose 0k}_n \nu_i 
{\bf k^-} {\bf i^+} I_n^{(d)}~.
 \label{six}
\end{equation}
Here the `deviations' $(\nu_k -1)$ of the indices from 1 are explicitly
separated so that all indices $\nu_k = 1$ do not contribute in the second
sum in curly brackets on the r.h.s.
Finally the double sum can be completely reduced to a single sum by means of
\cite{Andrey,ovt1}
\begin{equation}
\sum_{j=1}^n \nu_j {\bf j^+}I^{(d+2)}_n~ = - I^{(d)}_n~.
  \label{extra}
\end{equation}
This relation reduces simultaneously indices and dimension, which is what one 
wants
in general. It is not possible, however, to introduce it directly into (\ref{six})
since in (\ref{six}) we explicitly have to separate the term $i=k$.
Therefore, further details of how to apply  recurrence relation (\ref{recurseJ})
can be given only in the case of explicit examples (see Sect. 3).
In our notation recurrence relation (\ref{reduceJandDtod}) now reads
\begin{equation}
\left(  \right)_n
 \nu_j{\bf j^+} I^{(d+2)}_n=
\left[  - {j \choose 0}_n +\sum_{k=1}^{n} {j \choose k}_n
 {\bf k^-} \right] I^{(d)}_n,
 \label{eight}
\end{equation}
and recurrence relation (\ref{reduceDtod})
\begin{equation}
  (d-\sum_{i=1}^{n}\nu_i+1) \left(  \right)_n  I^{(d+2)}_n=
  \left[ {0 \choose 0}_n - \sum_{k=1}^n {0 \choose k}_n {\bf k^-} \right]I^{(d)}_n.
  \label{nine}
\end{equation}
Relations (\ref{six}) (including (\ref{extra})), (\ref{eight}) and (\ref{nine})
now replace (\ref{recurseJ}),(\ref{reduceJandDtod}) and (\ref{reduceDtod}).
For $\nu_i=1$ (\ref{eight}) and (\ref{nine}) correspond to eqs. (20) and (18)
of the first paper of \cite{BDK}.
In Sect. 3 we demonstrate how to reduce tensor integrals to scalar ones,
to which these recurrence relations are then applied.

Furthermore the following general observation is needed in what follows.
Applying the recursion relations, contractions of the $n$-th line occur. In this
case we encounter integrals with $p_n \neq 0$. In order to apply our recursion
relations for this case, we must properly define $G_{n-1}$. We can either shift
all momenta by $p_n$ or we can use from the very beginning the definition of
$G_{n-1}$ in terms of (\ref{MCD}) with $p_n \neq 0$. In this manner
the recurrence relations remain unchanged. 

   If our recurrence relations are to be implemented in terms of a
computer-algebra program, the following steps are recommended. At first, by
using (\ref{tensint}) we reduce tensor integrals to scalar ones with shifted
dimension (see also below). In a second step we apply (\ref{reduceJandDtod},\ref{eight})
and then (\ref{reduceDtod},\ref{nine}). Both relations produce the same factor $G_{n-1}$
in the denominator and therefore one expects some similarity of the obtained expressions.
Finally, relations (\ref{recurseJ},\ref{six}) are applied, bringing all integrals to a set of
master integrals in the generic dimension with powers 1 of the scalar propagators.

   The properties mentioned in the introduction leading to simplifications
for $n \geq 5$, which will be explained in more details below, are difficult
to be implemented in computer algebra systems. Therefore we prefer specific
representations, e.g. to avoid Gram determinants \footnote {Strictly speaking,
here we only mean $G_{n-1}$.}. In particular, to
obtain results as compact as possible, we will do a careful analysis
of the recurrence relations and different properties of Gram determinants in Sect. 3.

\subsection{Integrals with zero kinematic determinants}


When one or both of the determinants $G_{n-1}$ and $\Delta_n$ are equal to zero
the reduction procedure must be modified. 
In such cases it is possible to express integrals with
$n$ lines as a combination of integrals with $n-1$ lines.

First let us consider $G_{n-1}=0$. This is the case for $n$-point functions
with $n \geq 6$ for external 4-dimensional vectors since then the
order of $()_n$ is $n+1$ but its rank is 6.
In this case the recurrence relations  (\ref{eight})
and (\ref{nine}) cannot be directly used. Prior
to the reduction of $d$ one can remove one of the lines of the
diagram by using a relation which follows from (\ref{reduceJandDtod}):
\begin{equation}
  I^{(d)}_n= -\sum_{k=1}^{n}\frac{ (\partial_j \partial_k \Delta_n)}
  {(\partial_j  \Delta_n)}~ {\bf k^-}  I^{(d)}_n.
  \label{Gnzero}
\end{equation}
Here we keep the original form in terms of derivatives w.r.t. the masses since
in some computer algebra systems these derivatives may be easier to
calculate than determinants.

By repeated application of this relation one of the lines will
be contracted.  The procedure must be repeated until one 
obtains integrals with non-vanishing determinant $G$. After that
the reduction of the space-time dimension can be done for integrals
with the smaller number of lines.  
For $G_{n-1}=0$ yet another relation can be obtained from  
(\ref{reduceDtod}):
\begin{equation}
  I^{(d)}_n= -\sum_{k=1}^{n}\frac{ ( \partial_k \Delta_n ) }
  { 2 \Delta_n}~ {\bf k^-}  I^{(d)}_n.
  \label{Gnzero2}
\end{equation}
In fact both decompositions are equivalent since for $G_{n-1}=0$,
cancelling the common factor in $\Delta_n$ and 
$\partial_k \Delta_n$, the following relation holds
\begin{equation}
  \frac{  \partial_k \Delta_n} { 2 \Delta_n}=
  \frac{  \partial_k \partial_i \Delta_n}{\partial_i \Delta_n}.
  \label{puzzle}
\end{equation}
We will see in the next Sect. that indeed further important consequences 
follow from $G_{n-1}=0$.\\

Zeros of $\Delta_n$ may occur in special domains of the phase space for n=6.
In this case the reduction procedure will be as follows.
First, applying the relation
\begin{equation}
  (d-\sum_{i=1}^{n}\nu_i-1)G_{n-1}I^{(d)}_n=
  \sum_{k=1}^n (\partial_k \Delta_n) {\bf k^-} I^{(d-2)}_n,
  \label{DelZero}
\end{equation} 
which follows from (\ref{reduceDtod}), the integral with $n$ 
propagators must be reduced to a sum of integrals with $n-1$ 
propagators but in lower space-time dimension. As in
the case with $G_{n-1}=0$ the procedure of contracting lines
can be repeated until we obtain integrals with 
non-vanishing $\Delta_n$. After that integrals with 
space-time dimension $d < 4$ can be expressed in terms of integrals
in the generic dimension $d$ by using (\ref{extra})
and indices can be reduced to 1 by using relations
(\ref{recurseJ}) and (\ref{reduceJandDtod}).\\

   Explicit calculation \footnote {This has been performed for
$n=7$ and $8$ in terms of a FORM program, using component representations
of fourvectors.} shows that $\Delta_n=0$ 
for $n$-point functions
with $n \geq 7$, again for external 4-dimensional vectors - and
in particular for arbitrary masses. Then, due to (\ref{arbam})
also ${0\choose j}_n=0$ for $n \geq 7$. These properties eliminate
quite many terms in relations (\ref{six}), (\ref{eight}) and (\ref{nine})
\footnote {See also the discussion after (\ref{tausend}).}.
This is the reason why the investigation of $n \geq 7$ is postponed
at present. Nevertheless, we write the recurrence relations in what follows
for arbitrary $n$. In particular for external vectors not of dimension
$d=4$ the Gram determinants do not vanish as described and for these
cases anyway our relations hold in the given form for arbitrary $n$.


\section{Explicit recursions}

In this section we give details of how to calculate in particular 5- and 
6- point functions since these are of greatest actuality for present day experiments.
Our goal is to present the most compact formulae for direct applications.
Throughout we use the notation
\begin{equation}
  \int^d \equiv \int \frac{d^d q}{\pi^{d/2}}
\end{equation}
and for the scalar integrals we introduce the explicit notation
\bigskip
\noindent
\begin{equation}
  I_{p,ijk \ldots}^{[d+]^l,stu\ldots} = \int^{[d+]^l} \prod^n_{r=1}
  c_r^{-(1+\delta_{ri} +
    \delta_{rj} + \delta_{rk} + \ldots - \delta_{rs} - \delta_{rt} - 
    \delta_{ru} -
    \ldots)}, 
  \label{eins}
\end{equation}
where $ [d+]^l = 4 + 2l - 2\varepsilon $ . Observe  that ( number of
entries $ s,t,u,\ldots \neq i,j,k,\ldots ) + p = n $ and equal upper and lower indices `cancel'. 
The index $p$
specifies the `actual' number of external legs. We consider now 
Feynman diagrams in
the `generic' dimension $d = 4 - 2 \varepsilon$. For scalar $n$-point 
integrals in
the generic dimension we use the notation $I_n$.\\

   For the reduction of the tensor integrals to scalar ones with shifted dimension,
we wrote a FORM \cite{FORM}  program which applies (\ref{tensint})
to integrals of rank 1, 2 and 3. After inspection the result reads
for arbitrary $n$ and powers of the scalar propagators equal 1 (the latter
is the most frequent case in the electroweak Standard Model in the Feynman gauge; 
otherwise, as mentioned above, we can reduce higher indices by recurrence relations):


\begin{eqnarray}
  I_n^\mu &=& \int^d q^\mu \prod^n_{r=1} c^{-1}_r \nonumber\\
  &=& \sum^{n-1}_{i=1} p^\mu_i I_{n,i}^{[d+]} \label{drei}\\
  I_n^{\mu\nu} &=& \int^d q^\mu q^\nu \prod^n_{r=1} c^{-1}_r 
  \nonumber\\
  &=& \sum^{n-1}_{i,j=1} p_i^\mu p_j^\nu \cdot n_{ij} I_{n,ij}^{[d+]^2} 
  - \frac{1}{2} g^{\mu\nu} I_n^{[d+]}
  \label{vier}\\
  I_n^{\mu\nu\lambda} &=& \int^d q^\mu q^\nu q^\lambda \prod^n_{r=1} 
  c^{-1}_r
  \nonumber\\ &=& - \sum^{n-1}_{i,j,k=1} p_i^\mu p_j^\nu p^\lambda_k 
  \cdot n_{ijk}!
  I_{n,ijk}^{[d+]^3}\nonumber\\ & & + \frac{1}{2}  \sum^{n-1}_{i=1} 
  (g^{\mu\nu}
  p_i^\lambda + g^{\mu\lambda} p_i^\nu + g^{\nu\lambda} p_i^\mu) 
  I_{n,i}^{[d+]^2},
  \label{fuenf}
  \end{eqnarray}
where $ n_{ij} = 1 + \delta_{ij}$ and
$ n_{ijk} = 1+\delta_{ij} + \delta_{ik} +  \delta_{jk} -
  \delta_{ij} \delta_{ik} \delta_{jk} $
  is the number of equal indices among $i,j,k$ which can be
  written in this symmetric manner.\\

  We first consider now $n=5$, using recursion relation (\ref{eight})
  we have:
  \begin{equation}
    {\nu}_{ijk}~I_{5,ijk}^{[d+]^3} =  - \frac{{0\choose k}_5}{{ \choose }_5}
    I_{5,ij}^{[d+]^2} + \sum^5_{s=1} \frac{{s\choose k}}{{~\choose ~}_5}
    I_{p,ij}^{[d+]^2,s}~,~~~~~{\nu}_{ijk} = 1+\delta_{ik}+\delta_{jk} \label{sechs} 
  \end{equation} 
  where in the second term on the r.h.s. we have introduced the index 
  $p$. In general
  $p=4$, but for $s=i,j$ we have $p=5$ (recall that equal upper and 
  lower indices cancel). Let us now consider the first term on the
  r.h.s. of (\ref{sechs}): 
  \begin{eqnarray} 
    {\nu}_{ij}~I_{5,ij}^{[d+]^2} &=&  - \frac{{0
        \choose j}_5}{{~\choose ~}_5} I_{5,i}^{[d+]} + \sum^5_{s=1} \frac{{s 
        \choose
        j}_5}{{~\choose ~}_5} I_{p,i}^{[d+],s}~,~~~~~{\nu}_{ij}=1+{\delta}_{ij} \label{acht}\\
    I_{5,i}^{[d+]} &=& - \frac{{0 \choose i}_5}{{~ \choose ~}_5} I_5 + 
    \sum^5_{s=1}
    \frac{{s\choose i}}{{~ \choose ~}_5} I^s_4\label{neun}\\ 
    I_{p,i}^{[d+],s} &=& -
    \frac{{0 s\choose i s}_5}{{s \choose s}_5}I_4^{s}+\sum^5_{t=1} 
    \frac{{t s\choose i s}_5}{{s\choose s}_5}  I_3^{st}
    ~~~~~~~(s \neq i,~ p=4), \label{zehn} 
  \end{eqnarray} 
  where we indicated two more steps of
  the reduction. Like in (\ref{sechs}) we have again introduced
  the index $p$ in the second term on the r.h.s. of (\ref{acht}): for 
  $s=i,~p=5$. 
  At this point it is also worth pointing out that
  \begin{equation}
    n_{ij}=\nu_{ij} ,~~~~~ n_{ijk}!=\nu_{ij} \nu_{ijk}.
    \label{sieben}
  \end{equation}
  Now consider the second term on the r.h.s. of (\ref{sechs}): 
  \begin{equation}
    {\nu}_{ij}~I_{p,ij}^{[d+]^2,s} =  - \frac{{0 s\choose j s}_5}{{s \choose 
        s}_5} I_{4,i}^{[d+],s} + \sum^5_{t=1}
    \frac{{t s\choose j s}_5}{{s\choose s}_5}
    I_{q,i}^{[d+],st}  ~~~~~~~(s \neq i,j, ~p=4). 
    \label{elf}
  \end{equation} 
  Here the index $q$ in the second term on the r.h.s. is $q=3$ in
  general, except when $t=i$, in which case $q=4$ (see 
  (\ref{fuenfzehn})). For $q=3$
  we have: 
  \begin{equation} 
    I_{3,i}^{[d+],st} = - \frac{{0 s t\choose i s  t}_5}{{s t\choose s 
        t}_5} I_3^{st} + \sum^5_{u=1} \frac{{u s t\choose i s t}_5}{{s 
        t\choose s t}_5} I_2^{stu}\ .\label{zwoelf}
  \end{equation}
  For $s=i,j$ in (\ref{elf}) and also in the second part of 
  (\ref{fuenf}) we have
  integrals of the type 
  \begin{equation} I_{5,i}^{[d+]^2} = - \frac{{0 \choose
        i}_5}{{~ \choose ~}_5} I_5^{[d+]} + \sum^5_{s=1} \frac{{s\choose i\ 
        }_5}{{~ \choose
        ~}_5} I_4^{[d+],s} \label{dreizehn} 
  \end{equation} 
  and finally, applying recursion (\ref{reduceDtod}), we have 
  \begin{eqnarray} 
    I_5^{[d+]} &=& \left[ \frac{{0 \choose 0}_5}{{~
        \choose ~}_5} I_5 - \sum^5_{s=1} \frac{{0 \choose s}_5}{{~ \choose 
        ~}_5}
    I_4^s\right] \cdot \frac{1}{d-4} \label{vierzehn}\\ 
    I_4^{[d+],s} &=& \left[
    \frac{{0 s\choose 0 s}_5}{{s\choose s}_5}
    I_4^s - \sum^5_{t=1} \frac{{0 s\choose t s}_5}{{ s\choose s}_5} 
    I_3^{st} \right] \cdot \frac{1}{d-3}.
    \label{fuenfzehn} 
  \end{eqnarray} 
  We observe that (\ref{vierzehn}) is unpleasant in the sense that the 
  expression in
  the square bracket is zero for $d=4$ and the overall factor is 
  $-\frac{1}{2 \varepsilon}$, i.e.
  we need to expand the 5- and 4- point functions up to order 
  $\varepsilon$ in order
  to get the finite part. In fact, however, we will show that
  $I_5^{[d+]}$ cancels. To demonstrate this for 
  the 5-point function,
  we need first of all to express the $g^{\mu \nu}$ tensor in terms of the
  external vectors.
  Assuming that the external vectors $p_1, \ldots , p_4$ are 4-dimensional
  and independent 
  (i.e. no collinearities occur), we can write ( see e.g. \cite{Schouten}  )
  \begin{equation}
    g^{\mu\nu} = 2 \sum^4_{i,j=1} \frac{{i\choose j}_5}{{~\choose ~}_5} 
    p_i^\mu p_j^\nu \label{siebzehn} 
  \end{equation} 
  Inserting this into (\ref{vier}), we get
  \begin{equation} 
    \sum^4_{i,j=1} p_i^\mu p_j^\nu \cdot n_{ij} I_{5,ij}^{[d+]^2} -
    \sum^4_{i,j=1} p_i^\mu p_j^\nu \frac{{i\choose j}_5}{{~\choose ~}_5} 
    I_5^{[d+]}\ . \label{achtzehn} 
  \end{equation} 
  By inspection of (\ref{acht}) we see that the
  second term on the r.h.s. for $s=i$ exactly contains the $I_5^{[d+]}$ 
  to cancel the
  $I_5^{[d+]}$ in (\ref{achtzehn}). In a similar way with the help of
  (\ref{siebzehn}) we can rewrite the second sum in (\ref{fuenf}), to 
  get
  \begin{equation} 
    \sum^4_{i,j,k=1} p_i^\mu p_j^\nu p_k^\lambda  \left[
    \frac{{j\choose k}_5}{{ \choose }_5} I_{5,i}^{[d+]^2} + 
    \frac{{i\choose k}_5}{{
        \choose }_5} I_{5,j}^{[d+]^2} + \frac{{i\choose j}_5}{{ \choose }_5}
    I_{5,k}^{[d+]^2} \right]. \label{neunzehn} 
  \end{equation} 
  To pick out the $I_5^{[d+]}$ we use (\ref{dreizehn}).
  Inserting (\ref{sechs}) into
  (\ref{fuenf}), the $I_5^{[d+]}$ contributions come from the second 
  term in
  (\ref{acht}) (for $s=i$) and the second term of (\ref{sechs}) (for 
  $s=i,j$).
  Further we need the property (\ref{sieben}).

  Finally a remark is in order concerning the finiteness of the 5-point 
  function: the
  only infinities occurring in the above decomposition are coming from 
  the $I_2~'s$
  in (\ref{zwoelf}). The $\frac{1}{\varepsilon}$ terms in these 2-point 
  functions are
  independent of masses and momenta, however. Thus according to (see \cite{melrose})
  \begin{equation} 
    \sum^n_{j=1} {i \choose j}_n  =  ~0~~ ,~~~i = 1, \cdots  n
    \label{zweiundzwanzig}
  \end{equation} 
  for $n=3$ these infinite terms cancel. With these considerations the tensor 5-point
functions can be completely reduced to scalar 4-, 3- and 2-point functions in generic
dimension. We want to point out that use has been made only of relations
(\ref{eight}) and (\ref{nine}). We will see below that (\ref{neun}) should be 
replaced by another relation containing no Gram determinant, which might simplify
the numerics in some kinematic domains. At the end
the tensor integrals under consideration are finite (after cancellation of the
above mentioned terms of order $\frac{1}{\varepsilon}$ ) and therefore we can put
$\varepsilon = 0$, which in particular applies to the scalar 5-point function 
\cite{brown,melrose,BDK}.\\

  For $n=6$ and $d=4$ the situation is completely different due to the fact that 
  $()_6=0$ as was discussed in Sect. 2.3~.
  Therefore (\ref{reduceJandDtod}) and (\ref{reduceDtod})  both reduce to
  (\ref{Gnzero}), i.e. instead of those two recursion relations we have only one. 
  Explicitly it reads, introducing the signed minors of the modified Cayley 
  determinant
  \begin{equation} 
    I_{6,i..}^{[d+]^l} = \sum^6_{r=1} \frac{{R\choose
        r}_6}{{R\choose 0}_6} I_{p,i..}^{[d+]^l,r}, 
~~~R~=~\mathrm{any}~~\mathrm{value}~~ 0, 
    \ldots ,6. \label{zwanzig} 
  \end{equation} 
  We see that (\ref{zwanzig}) does not allow to reduce
  the dimension and it can only be used to reduce indices! 
  If $p=5$ on the r.h.s. of (\ref{zwanzig}), which
  is the case for $r \neq i..$,
  then the reduction can be continued as for the 5-point function 
  above. If, however,
  $p=6$ on the r.h.s. of (\ref{zwanzig}) in case of $r = i..$, then again 
  (\ref{zwanzig}) has to be applied until all $I_{6,i..}^{[d+]^l}$ are 
  eliminated.\\

  So far only relations (\ref{eight}) and (\ref{nine}) for the Gram
  determinants $G_n$ have been applied.  Now let us investigate
  (\ref{six}). It does not contain Gram determinants $G_n$. A priori it does
  not allow to reduce the dimension.  In the following, however, we
  will explicitly show that (\ref{extra}) can be used to perform one
  sub-summation in (\ref{six}) and that it is this relation which
  finally provides a very efficient possibility to reduce also the
  dimension in the case of the 6-point function. Also some further
  useful information about the reduction of 5-point functions is
  obtained in this manner. In the following we investigate explicitly
  the integrals occurring in (\ref{drei}),(\ref{vier}) and
  (\ref{fuenf}).

    For arbitrary $n$ (\ref{six}) yields for the integral in (\ref{drei})
    \begin{equation}
      {0\choose 0}_n I^{[d+]^l}_{n,i} = \left[ n+1 - (d+2l)\right] 
      {0\choose i}_n I_n^{[d+]^l}
      - \sum^n_{r, s\atop r\not= s}  {0i\choose 0r}_n I_{n-1,s}^{[d+]^l
        ,r}\label{fuenfundzwanzig} 
    \end{equation} 
    and using (\ref{extra}) once we
    obtain  : 
    \begin{equation} 
      {0\choose 0}_n  I_{n,i}^{[d+]^l} = \left[ n+1 -(d+2l) \right]
      {0\choose i}_n I_n^{[d+]^l} + \sum^n_{r= 1} {0 i\choose 0 r}_n 
      I_{n-1}^{[d+]^{(l-1)},r}~.
      \label{sechsundzwanzig} 
    \end{equation} 
    Similarly, for the integral in (\ref{vier}) we obtain from
    (\ref{six}) for arbitrary $n$ 
\begin{eqnarray} 
{0\choose 0}_n {\nu}_{ij} I_{n,ij}^{[d+]^l} &=& \left[ 
n+2 - (d+2l)\right]
{0\choose j}_n I_{n,i}^{[d+]^l } - {0 j\choose 0 i}_n \sum^n_{s=1} 
I_{n,s}^{[d+]^l}
\nonumber\\ & & - \sum^n_{r=1\atop r\not= i} {0
j\choose 0 r}_n \sum^n_{s=1\atop s\not= r} {\nu}_{is} 
I_{n-1,is}^{[d+]^l,r}
\label{siebenundzwanzig}
\end{eqnarray}
and using (\ref{extra})
\begin{eqnarray}
{0\choose 0}_n {\nu}_{ij} I_{n,ij}^{[d+]^l} &=& \left[ n+2-(d+2l)\right] 
{0\choose j}_n
I_{n,i}^{[d+]^l } + {0 i\choose 0 j}_n I_n^{[d+]^{(l-1)}}\nonumber\\ 
& & +
\sum^n_{r=1\atop r\not= i} {0 j\choose 0 r}_n 
I_{n-1,i}^{[d+]^{(l-1)},r} ,
\label{achtundzwanzig} 
\end{eqnarray} 
It is interesting to note that in
(\ref{sechsundzwanzig}) for $n=5$ and $l=1$ as well as in 
(\ref{achtundzwanzig}) for
$n=6$ and $l=2$, the numerical square brackets evaluate to $4-d=2 
\varepsilon$. This
means a great simplification for $\varepsilon=0$ and in particular
(\ref{sechsundzwanzig}) ought to replace (\ref{neun}) for $n=5$
( a little algebra shows that indeed for $n=5$ (\ref{sechsundzwanzig})
and (\ref{neun}) are identical including the $\varepsilon$-part). 
Moreover this is
exactly what is needed in (\ref{achtundzwanzig}) for $n=6$ and $l=2$, 
the integral needed in (\ref{vier}) for $n=6$. 


For the integral with three indices occurring in (\ref{fuenf}) 
we finally have 
\begin{eqnarray}
{0\choose 0}_n {\nu}_{ijk} I_{n,ijk}^{[d+]^l} &=& \left[ n+3-(d+2l)\right] 
{0\choose k}_n I_{n,ij}^{[d+]^l }  \nonumber\\
& &+\left[ {0 k\choose 0 i}_n I_{n,j}^{[d+]^{(l-1)}} +
{0 k\choose 0 j}_n I_{n,i}^{[d+]^{(l-1)}} \right]\frac{1}{\nu_{ij}}+
\sum^n_{r=1\atop r\not= i,j} {0 k\choose 0 r}_n
I_{n-1,ij}^{[d+]^{(l-1)},r}  .~~~~~~~~
\label{tausend} 
\end{eqnarray}
Equations (\ref{sechsundzwanzig}),(\ref{achtundzwanzig}) and (\ref{tausend})
are valid for any $n$ and they represent in a way an optimal form 
of the recursion for the considered integrals. For $n \ge 7$ and $d=4$, as 
mentioned in Sect. 2.3 already, these relations are still valid but reduce
considerably. Since this case is at present of minor physical interest,
it will be discussed separately. We only mention that, e.g., using (\ref{tausend})
for $i=j=k$ and $n=7$ we obtain a relation for $I_{7,i}^{[d+]^{l}}$
in terms of $I_{6,ii}^{[d+]^l,r}$, summed over $r$, etc.

The integrals with highest 
dimension in the above relations have coefficients ${0\choose i}_n$,
${0\choose j}_n$ and ${0\choose k}_n$, respectively. In (\ref{drei}), (\ref{vier})
and (\ref{fuenf}) these are multiplied with $p_i^{\mu}$, $p_j^{\nu}$ and $p_k^{\lambda}$
and summed over $i,j,k$. Due to 
\begin{equation}
\sum^{n-1}_{j=1} p_j^{\nu} {0 \choose j}_n = 0~,~~~~~n \ge 6
\label{tausendeins}
\end{equation}
all these contributions vanish. To prove (\ref{tausendeins}), 
we project it
on all $p_i, i=1 \ldots n-1 (p_n=0)$ in $d$=4 dimensions, assuming that
at least four $p_i$'s are linearly independent. Thus we have to show
\begin{equation}
\sum^n_{j=1} p_i p_j {0\choose j}_n = 0 ~ , ~~~~~i=1 \ldots n-1.
\label{tausendzwei}
\end{equation}
First of all we write the scalar products in the form \footnote {This essential 
decomposition, to our knowledge,
was first used in \cite{melrose} and is the basis of the proof in \cite{BDK} as well.}
\begin{eqnarray}
p_i p_j = \frac{1}{2} \left\{ Y_{ij} - Y_{in} - Y_{jn} + Y_{nn} \right\} 
\end{eqnarray}
so that (\ref{tausendzwei}) reads
\begin{equation}
\sum^n_{j=1} \left\{ Y_{ij} - Y_{nj} - (Y_{in} - Y_{nn})\right\} {0 
\choose j}_n = 0 \ 
\label{tausenddrei}
\end{equation}
Now the term $(Y_{in} - Y_{nn})$ is independent of $j$ and the summation over
$j$ can therefore be performed:
\begin{equation}
\sum^n_{j=1} {0\choose j}_n = ()_n = 0~, ~~~~~ n \ge 6.
\end{equation}
Next we consider the first term in (\ref{tausenddrei})
\begin{equation}
\sum^n_{j=1}Y_{ij} {0\choose j}_n = \left\vert 
\matrix{0 & Y_{i1} & Y_{i2} & Y_{i3} & \ldots & Y_{in}\cr
        1 & Y_{11} & Y_{12} & Y_{13} & \ldots & Y_{1n}\cr
        1 & Y_{12} & Y_{22} & Y_{23} & \ldots & Y_{2n}\cr
   \vdots &        &        &        &        &       \cr
        1 & Y_{i1} & Y_{i2} & Y_{i3} & \ldots & Y_{in}\cr
   \vdots &        &        &        &        &       \cr
        1 & Y_{1n} & Y_{2n} & Y_{3n} & \ldots & Y_{nn}\cr}       
\right\vert  = - {0\choose 0}_n
\label{tausendvier}
\end{equation}
Subtracting the $0$-th line of this determinant from the $i$-th one,
the latter takes the form $(1,0, \ldots 0)$. Expanding finally the determinant
w.r.t. the first column yields the r.h.s. of (\ref{tausendvier}). Thus this determinant
is independent of $i$, which finally proves (\ref{tausenddrei}), and
hence (\ref{tausendeins}).

Further we also see that $I_n^{[d+]^{(l-1)}}$ in (\ref{achtundzwanzig})
cancels against the $g^{\mu \nu}$ term in (\ref{vier}) ($l=2$) due to
\begin{equation}
g^{\mu\nu} = 
\frac{2}{{0\choose 0}_n} \sum^{n-1}_{i,j=1} {0~i\choose 
0~j}_n p_i^\mu p_j^\nu~, ~~~~~ n \ge 6 .
\label{tausendfuenf}
\end{equation}
As above we show this again by proving that all projections of this tensor 
on any pair of vectors $p_k, p_l (k,l=1 \ldots n-1)$ vanish, i.e.
\begin{equation}
\sum^n_{i,j=1} (p_i p_k) (p_j p_\ell) {0~i\choose 0~j}_n = 
\frac{1}{2} p_k p_\ell {0\choose 0}_n .
\end{equation}
Expressing all scalar products again in terms of the $Y_{ik}$, this is
equivalent to
\begin{eqnarray}
& &\sum^n_{i,j=1} \left[ Y_{ik} - Y_{in} - (Y_{kn} -
Y_{nn})\right] \left[ Y_{j\ell} - Y_{jn} - (Y_{\ell n} - Y_{nn}
)\right] {0~i\choose 0~j}_n \nonumber \\
& &= \left[ Y_{k\ell} - Y_{kn} - (Y_{\ell n} - Y_{nn} )\right]
{0\choose 0}_n 
\label{tausendsechs}
\end{eqnarray}
The contribution on the l.h.s. of (\ref{tausendsechs}) with $Y$'s
independent of $i,j$ is proportional to 
\begin{equation}
\sum^n_{i=1} \sum^n_{j=1} {0~i\choose 0~j}_n = - \sum^n_{i=1} 
{0\choose i} = - {~\choose ~}_n = 0.
\end{equation}
The `linear' terms, according to (\ref{vierundzwanzig}) and 
(\ref{tausendvier}), read
\begin{equation}
\sum^n_{i=1} Y_{ik} \sum^n_{j=1} {0~i\choose 0~j} = - {0\choose 
0}_n
\end{equation}
and are independent of $k$, i.e. the contributions of this type cancel. 
Finally, the `quadratic' terms are of the form
\begin{equation}
\sum^n_{i=1} Y_{ik} \sum^n_{j=1} Y_{jl} {0~i\choose 0~j} = 
\sum^n_{i=1} Y_{ik} \cdot \delta_{il} {0\choose 0}_n = Y_{k\ell} 
{0\choose 0}_n,
\end{equation}
which proves (\ref{tausendsechs}) and thus (\ref{tausendfuenf}).\\

We point out that (\ref{tausendfuenf}) is a representation of the 
$g_{\mu \nu}$ tensor in close analogy to (\ref{siebzehn}), only that
in this case the $n-1$ vectors ($n \ge 6$) are linearly dependent
in $d=4$ dimensions.


Finally we investigate the simplification of (\ref{fuenf}) due to 
(\ref{tausend}), where we concentrate on the 6-point function, $n=6$. 
Let us first consider the $I_{n,i}^{[d+]^2}$ integrals.
These are given by (\ref{sechsundzwanzig}) for $l=2$. First of all we
again observe that their first part can be dropped due to (\ref{tausendeins}).
Their remaining parts for $n=6$ are only of the $I_5^{[d+]}$ type, which
we would like to cancel. This concerns in particular the complete
second part in (\ref{fuenf}). It is easy to see that due to (\ref{tausendfuenf})
the $g^{\mu \lambda}$ and $g^{\nu \lambda}$ terms cancel against the
two terms in the square bracket of (\ref{tausend}) by summing in (\ref{fuenf}) 
over $i,k$ and $j,k$~, respectively. To show the cancellation 
of the remaining $I_5^{[d+]}$'s is a bit tricky. At first we observe
that the last term in (\ref{tausend}) contains the following $I_5^{[d+]}$ term
(most easily seen from (\ref{acht})):
\begin{equation}
\nu_{ij} I_{5,ij}^{[d+]^2,r} = \frac{{i r\choose j r}}{{r\choose r}} I_5^{[d+],r} 
+ \cdots ~~.
\end{equation} 
Thus the sum of all remaining $I_5^{[d+]}$ terms is given by
\begin{equation}
-{{{0 \choose 0 }}_6}^{-1} \sum^5_{i,j,k=1} p_i^\mu p_j^\nu p^\lambda_k 
\sum_{r=1}^6 \frac{{i r\choose j r}}{{r\choose r}} {0 k\choose 0 r} I_5^{[d+],r}
+\frac{1}{2} g^{\mu\nu} {{{0 \choose 0 }}_6}^{-1} 
 \sum^5_{k=1} p_k^\lambda \sum_{r=1}^6 {0 k\choose 0 r} I_5^{[d+],r} 
\label{finish}
\end{equation}
Then we observe that for $n=6$ the $g^{\mu \nu}$ tensor can be written in the 
following form
\begin{equation}
g^{\mu\nu} = 2 \sum^5_{i,j=1} \frac{{i r\choose j r}_6}{{r \choose r}_6} 
p_i^\mu p_j^\nu~, ~~~~~ r=1 \ldots 6.
\label{finishM10}
\end{equation} 
For $r \leq 5$ this equation immediately follows from (\ref{siebzehn}), just by
scratching one of the 5 momenta and assuming that always four of them are linearly
independent. For $r=6$, (\ref{finishM10}) can be proven in the same
way as (\ref{tausendfuenf}). 
Thus, summing over $i,j=1 \ldots 5$ in (\ref{finish}),
the cancellation of all $I_5^{[d+]}$ integrals is also shown for the tensor integrals
of rank 3, a result which has been obtained previously in \cite{BDK}. 
What remains in (\ref{tausend}) is finally again only the last sum with the
exclusion of the $I_5^{[d+]}$ contribution. The $I_{5,ij}^{[d+]^2,r}$ integrals in (\ref{tausend})
can be calculated, e.g., by means of (\ref{acht}), without the term $s=i$ in the sum
on its r.h.s. or from (\ref{achtundzwanzig}) again dropping $I_5^{[d+]}$ on the r.h.s.~.\\

   As a final remark we point out that the only Gram determinants occurring are those
in the tensor integrals of rank 3, coming from the reduction in (\ref{acht}), where
instead of (\ref{neun}) now (\ref{sechsundzwanzig}) is to be used.
Thus
only inverse Gram determinants of the first power appear. Moreover the remaining integrals
are all reduced to standard 4- and 5-point functions for $n=6$.

\bigskip
\noindent
{\Large{\bf Conclusion}}
\medskip

We have presented an algorithm for the calculation of $n$-point 
Feynman diagrams, applicable for any tensorial structure  and gave a 
detailed specification for 5- and 6-point functions. 
Dependent on the
kinematic situation different recurrence relations may be used.
A typical example is the equivalence of (\ref{neun}) and (\ref{sechsundzwanzig})
for $n=5$. 
All recursion relations are valid for arbitrary dimension.
In the main part of the paper we have concentrated on the experimentally most 
relevant cases, i.e. the tensor 5- and 6- point functions, and considered
$d=4$ for the external momenta in order to simplify the results. In this case
a regulator mass is needed for infrared divergences. We restricted ourselves
to tensors of rank 3 \footnote {For more
than three integration momenta in the numerator the extension of our procedure
is straightforward.}. 
For the 5-point function, due to the above equivalence,
there occur inverse Gram determinants of only second order; for the 6-point
function, surprisingly, only of first order. $n$-point functions with $n \geq 7$
are not considered explicitly since due to the drastic reduction of
the recurrence relations for $d=4$ these cases need a separate investigation.

\bigskip
\noindent
{\large{\bf Acknowledgment}}

\medskip
We are grateful to A. Davydychev and S. Dittmaier for useful discussions
and to A. D. also for carefully reading the manuscript. O.V. Tarasov
gratefully acknowledges financial support from the BMBF.

\newpage
{\Large{\bf Appendix}}
\medskip

In this appendix we present for the 6-point function the `effective' contributions
to the integrals (\ref{drei}) - (\ref{fuenf}), i.e. leaving out all those contributions
which we have shown to cancel. First of all no contributions come from the metric
tensor in (\ref{vier}) and (\ref{fuenf}). The integral in (\ref{drei}) is `effectively'
given by
\begin{equation}
I_{6,i}^{[d+]}:=\frac{1}{{0\choose 0}_6} \sum^6_{r=1} {{0 i\choose 0  r}_6} I_5^r = 
\frac{1}{{0\choose 0}_6} \sum^6_{r=1} \frac{{0 i\choose 0 r}_6}{{0 r\choose 0 r}_6}
\sum^6_{s=1} {{0 r\choose s r}_6} I_4^{rs},
\label{App1}
\end{equation}
where we have used (\ref{sechsundzwanzig}) and expressed the 5-point function
$I_5^r$ in generic dimension in terms of 4-point functions. No Gram determinant is
involved in this case. 

The remaining integral in (\ref{vier}) is `effectively'
\begin{equation}
n_{ij} I_{6,ij}^{[d+]^2}:=\frac{1}{{0\choose 0}_6} \sum^6_{r=1\atop r\not= i} {0 j\choose 0 r}_6
I_{5,i}^{[d+],r} :=\frac{1}{{0\choose 0}_6} \sum^6_{r=1} \frac{{0 j\choose 0 r}_6}
{{0 r\choose 0 r}_6} \sum^6_{s=1} {{0 i r\choose 0 s r}_6} I_4^{rs}.
\label{App2}
\end{equation}
Here (\ref{achtundzwanzig}) is used and the integral $I_{5,i}^{[d+],r}$ is expressed
in terms of (\ref{sechsundzwanzig}), where the square bracket of the first contribution
evaluates to $2 \varepsilon$ and is dropped. The result is very similar to (\ref{App1}),
i.e. it is a double sum over 4-point functions and no Gram determinant involved. The
latter property is due to (\ref{sechsundzwanzig}) , which is used instead of (\ref{neun}).

The integral in (\ref{fuenf}) is a bit more complicated:
\begin{eqnarray}
n_{ijk}! I_{6,ijk}^{[d+]^3}
&:=& \frac{1}{{0\choose 0}_6} \sum^6_{r=1\atop r\not= i,j} {0 k\choose 0 r}_6
\nu_{ij} I_{5,ij}^{[d+]^2,r} \nonumber \\
&:=& \frac{1}{{0\choose 0}_6} \sum^6_{r=1\atop r\not= i,j}
\frac{{0 k\choose 0 r}_6}{{0 r\choose 0 r}_6} \left\{ - {0 r\choose j r}_6 I_{5,i}^{[d+]^{2},r}
+ \sum^6_{s=1\atop s\not= i} {{0 j r\choose 0 s r}_6} I_{4,i}^{[d+],rs} \right\} .
\label{App3}
\end{eqnarray}
Here (\ref{tausend}) is used together with (\ref{achtundzwanzig}) for the reduction of $I_{5,ij}^{[d+]^2,r}$.
In the square bracket of the first part of (\ref{achtundzwanzig}) again $\varepsilon =0$ is
taken. Reducing $I_{5,i}^{[d+]^2,r}$ we see no possibility to avoid the inverse Gram determinant
and therefore propose to use (\ref{dreizehn}), i.e. `effectively' (dropping $I_5^{[d+]}$)
\begin{equation}
I_{5,i}^{[d+]^2,r}:=\frac{1}{{r\choose r}_6} \sum^6_{s=1} {s r\choose i r}_6 I_4^{[d+],rs}
\label{App4}
\end{equation}
with ($\varepsilon =0$)
\begin{equation}
I_4^{[d+],rs}=\frac{1}{{r s\choose r s}_6} \left[ {{0 r s\choose 0 r s}_6} I_4^{rs}
- \sum^6_{t=1} {{0 r s\choose t r s}_6} I_3^{rst} \right]
\label{App5}
\end{equation}
from (\ref{fuenfzehn}). Finally, $I_{4,i}^{[d+],rs}$ should be reduced to the form
(see e.g. (\ref{zehn}))
\begin{equation}
I_{4,i}^{[d+],rs}=\frac{-1}{{r s\choose r s}_6} \left[{{i r s\choose 0 r s}_6} I_4^{rs}
- \sum^6_{t=1} {{i r s\choose t r s}_6} I_3^{rst} \right]~.
\label{App6}
\end{equation}
The factor
${{r\choose r}_6}^{-1}$ in (\ref{App4}) is the only occurrence of the inverse Gram
determinant. ${{r s\choose r s}_6}^{-1}$ in (\ref{App5}) and (\ref{App6}) are in
principle inverse Gram determinants as well, but since one does not expect the l.h.s.
of (\ref{App5}) and (\ref{App6}) to have kinematic singularities for 
${{r s\choose r s}_6} \to 0$, the numerators also vanish in this case, which allows to
keep the numerical evaluation under control. The situation is different in (\ref{App4}) :
even if the l.h.s. has no kinematic singularity as ${{r\choose r}_6} \to 0$, on the
r.h.s. $I_5^{[d+]}$ has been cancelled and therefore the above argument is not applicable.

It is convenient to add up all contributions in (\ref{App3}) by introducing
the quantity
\begin{equation}
\left\{ r \right\}_{ij}^{st} ={0 r\choose 0 r}_6 \left[ {t r\choose s r}_6 {i r s\choose j r s}_6
-{r s\choose r s}_6 {i r s\choose j r t}_6 \right] - {r s\choose r s}_6 \left[ {t r\choose 0 r}_6
{i r 0\choose j r s}_6 - {s r\choose 0 r}_6 {i r 0\choose j r t}_6 \right].
\label{App7}
\end{equation}
With this definition (\ref{App3}) can be written in the form
\begin{equation}
n_{ijk}! I_{6,ijk}^{[d+]^3}:=\frac{1}{{0\choose 0}_6} \sum^6_{r=1\atop r\not= i,j}
\frac{{0 k\choose 0 r}_6}{{0 r\choose 0 r}_6} \frac{1}{{r\choose r}_6}\sum^6_{s=1\atop s\not=r}
\frac{1}{{r s\choose r s}_6} \left\{ -\left\{ r \right\}_{ij}^{s0} I_4^{rs}
+\sum^6_{t=1} \left\{ r \right\}_{ij}^{st} I_3^{rst} \right\}.
\end{equation}
As we see, some further cancellations occur. The representations
derived here are particularly useful for numerical evaluation.

\end{document}